\documentclass[
	a4paper, % Paper size, use either a4paper or letterpaper
	10pt, % Default font size, can also use 11pt or 12pt, although this is not recommended
	twoside, % Two side traditional mode where headers and footers change between odd and even pages, comment this option to make them fixed
]{LTJournalArticle}

\usepackage{etoolbox}
\usepackage{amsmath,amsfonts,stmaryrd,amssymb, enumitem} % Math packages

\usepackage{istgame}
\usepackage{csquotes}
\usepackage{graphicx,apacite}
\usepackage[ruled]{algorithm2e} % Algorithms
\usepackage{natbib}
\usepackage{pgfplots}
\usepackage[british]{babel} % Use British English
\usepackage{float}

\usepackage[framemethod=tikz]{mdframed} % Allows defining custom boxed/framed environments

\usepackage{listings} % File listings, with syntax highlighting
\newtheorem{remark}{Remark}
\newtheorem{definition}{Definition}
\providecommand{\keywords}[1]{\textit{Keywords: } #1}

\title{My Statistics is Better than Yours} % Article title, use manual lines breaks (\\) to beautify the layout

\author{Simon Benhaiem}
\date{\today}

\renewcommand{\maketitlehookd}{
	    \begin{abstract}
              Statistical schools---such as Bayesianism and Frequentism---are often presented as competing frameworks, each claiming technical rigour and superiority. Frequentism emphasizes objective inferences through repeated sampling, while Bayesianism incorporates prior beliefs and updates them with new evidence. Despite their strengths, neither school proves universally applicable, and the pursuit of a single “correct” statistical framework is ultimately misguided. Instead, this essay advocates for a context-dependent approach to statistical norms, drawing on \cite{douglas2004irreducible}'s concept of “operational objectivity”.
              The idea is that by aligning the context of the research question with the value judgments inherent to its field, a certain statistical paradigm is warranted. This essay explores the decision-theoretic foundations of Bayesianism, examines its descriptive limitations as highlighted by the Ellsberg paradox, and addresses the challenges of comparing different normative systems.
	    \end{abstract}
            \keywords{operational objectivity, normative theory for statistics, Bayesianism, Dutch book arguments.}
            
}

\begin{document}

\maketitle % Output the title section

%----------------------------------------------------------------------------------------
%	ARTICLE CONTENTS
%----------------------------------------------------------------------------------------
    \section{Introduction}
        When performing data analysis, a researcher often faces a choice between Frequentist and Bayesian approaches\footnote{There are other schools of statistics, such as Likelihoodism and Fiducial probabilities which will be discussed later in the essay. The Frequentist versus Bayesian dilemma is introduced here as it is one of the most active debates in statistics. See the Jeffreys-Lindley's paradox.}, each of which offers distinct principles and prescribed methods. Frequentism operates under the assumption of repeated sampling, aiming for so-called objective inferences through significance tests and efficient estimators. Bayesianism, on the other hand, incorporates a researcher’s prior beliefs\footnote{Strictly for \textit{subjective} Bayesianism.} about a hypothesis and updates them with new evidence to generate posterior distributions. Despite the technical rigour behind both methods, neither approach appears universally applicable. A single, “correct” statistical school may seem like an objective ideal. However, we will see that it becomes impossible to choose between the two schools, even when we try our best to fulfil this ideal. \\
        
        Instead, this essay proposes a \textit{context-dependent approach} to guide the selection of an appropriate statistical school. This type of approach is not novel. \cite{worsdale2021my}'s \textit{My Objectivity is Better than Yours} presents \cite{douglas2004irreducible}'s “operational” objectivity in the search for an objective gender inequality index. The authors point out the worrying obsession researchers have for finding a true measure of gender inequality that is universally applicable. Rather, \cite{worsdale2021my} recommend taking the research goals and context into “objectivity”, resulting in a context-dependent objectivity. We exploit the same idea and apply it to the search for a normative system of statistics, i.e. contextualizing statistical norms. \\

        The remainder of the essay is structured as follows. In Section \ref{sec: subjective Bayesianism} we introduce (subjective) Bayesianism and its decision-theoretic foundations. We do this specifically for Bayesian epistemology as its principles are quite simple and defended using compelling arguments. Frequentism and other schools have more ambiguous roots\footnote{However, the applications of Frequentism are well-defined and well-studied. In contrast, Bayesian applications are usually quite muddied, as it is not always clear what exact subschool the researcher is going for.}, which would warrant an essay of its own.  Section \ref{sec: Paradox} sketches a descriptive falsification of Bayesianism by the Ellsberg paradox. The problem is that empirical falsification is not enough to refute Bayesianism, as it is a normative idea. This, in turn, leads to the main issue of the paper: that there is no straightforward way to distinguish between two normative schools of statistics. Section \ref{sec: Meta-problem} proposes the two candidates: (i) the universalist approach and (ii) the context-dependent approach, and argues why the first one falls short. A small case study is also presented to give the essay some grit. Finally, we conclude in Section \ref{sec: Conclusion}. The Appendix can be found after the Biographical note.

        %Add subsection about the scope of the argument. I am arguing about the normative approach for choosing a normative system (meta-normativity). I invite discussions about the normative approach. 
    \section{Subjective Bayesianism}
    \label{sec: subjective Bayesianism}
        Bayesian epistemology tries to model inductive (fuzzy) logic using the \textit{credences} one has about a given hypothesis \citep{sep-epistemology-bayesian}. In particular, its axiomatic system controls how one ought to form initial credences (priors) and how to update them in light of new evidence. Accepting such a model as the normative system for statistics leads to Bayesian statistics, which uses these principles. Let us look at the core of these principles. \\

        We introduce some terminology. Let $\Omega$ be the sample space, i.e. the set of all possible outcomes, let $\mathcal{F}$ be the event space, a $\sigma$-algebra on $\Omega$. Then, the \textit{credence function} $\textnormal{Cr}: \mathcal{F} \to \mathbb{R}$ assigns a real number to every possible event. There are two principles a credence function typically adheres to:
        \begin{definition}[Probabilism]
            \textit{Probabilism} is a set of constraints on credences. It coincides with the definition of a probability measure. Namely, we need
            \begin{enumerate}
                \item $\textnormal{Cr}(A) \geq 0 \quad \forall \ A\in\mathcal{F}$,
                \item $\textnormal{Cr}(\Omega) = 1$,
                \item $\textnormal{Cr}(A \cup B)   =   \textnormal{Cr}(A) + \textnormal{Cr}(B) \: \textnormal{ for } \: A, B \! \! \in \mathcal{F} \: \textnormal{  where } A\cap B = \emptyset$.\footnote{The third constraint usually is generalized to countable collections $\{E_k\}_{k=1}^{\infty}$ of pairwise disjoint sets in $\mathcal{F}$.}
            \end{enumerate}
        \end{definition}
        \begin{definition}[Principle of Conditionalization]
            With a new piece of evidence $E\in \mathcal{F}$, one ought to change their credences according to the Principle of Conditionalization. Namely, for $\textnormal{Cr}(E) \neq 0$ and proposition $A\in\mathcal{F}$ our new credence is
            $$
            \textnormal{Cr}(A\mid E) = \frac{\textnormal{Cr}(A \cap E)}{\textnormal{Cr}(E)} .
            $$
        \end{definition}
        There are various debates on whether these definitions should be seen as normative constraints or as constructions. See \cite{sep-epistemology-bayesian} for a comprehensive survey on the objections.
        \\
        Until now, I have only described principles that govern how credences should be assigned and conditioned. Now we turn to the classic problem: how ought one form initial credences (priors)? There are two popular norms of prior choice: subjective and objective. 
        \begin{remark}[Prior choice]
            We have that
            \begin{enumerate}
                \item Subjective: every prior is allowed as long as it adheres to Probabilism.
                \item Objective: every prior is allowed as long as it adheres to Probabilism and is uninformative.
            \end{enumerate}
        \end{remark}
        This essay focuses explicitly on subjective Bayesianism. 
        
    \subsection{Foundation of subjective Bayesianism}
    \label{subsec: foundations}
        So how can the principles of subjective Bayesianism be defended? \cite{de1931sul} notoriously uses so-called Dutch Book Arguments (DBA):
        \begin{definition}[Dutch Book Arguments (DBA)]
            Dutch Book Arguments can be thought of as a device to identify credences that are irrational. In particular, a gambler's set of credences can result in a Dutch Book if a bookmaker can construct a series of bets that guarantees a net loss for the gambler. A rational agent is one that can never be Dutch-Booked. \\
            If, for example, Bayesianism offers a government for credences that is immune to Dutch Books, Bayesianism is a normative/rational model. Note that there may be multiple schools that are immune to Dutch Books. This means DBA are only a necessary condition for normativity/rationality.  
        \end{definition}
        As agents who use the principles of (subjective) Bayesianism cannot be \textit{Dutch-Booked}, they are rational decision-makers. de Finetti uses this device to vouch for (subjective) Bayesianism as a normative school of statistics, with the idea being that statistics essentially amounts to a data-dependent decision. \\
        Savage also backs (subjective) Bayesianism. In \cite{savage1972foundations}, he submits a set of postulates which are defended by presenting examples that seem irrefutable, the most popular being the \textit{sure thing principle}\footnote{Note that this is what Savage uses to adumbrate the sure thing principle, while the actual postulate is a formal statement.}:
        \blockquote{\textit{“A businessman contemplates buying a certain piece of property. He considers the outcome of the next presidential election relevant. [...] He asks whether he would buy if he knew that the Democratic candidate were going to win [...] and decides that he would. Similarly, he considers whether he would buy if he knew that the Republican candidate were going to win, and again finds that he would. Seeing that he would buy in either event, he decides that he should buy, even though he does not know which event obtains [...].”} (ellipses added)\\
        \cite{savage1972foundations}, p. 21}
        Despite the fact that Savage composed the postulate as a dominance principle, \cite{jeffrey1982sure} shows that it can be framed as a probabilistic prescription. \\
        
        %\footnote{Savage's second postulate also implies Likelihoodism; another statistical school that works in junction with subjective Bayesianism. See \cite{berger1988likelihood} for more detail.} 
        
        Here, it is ultimately important to recognize that both de Finetti’s and Savage’s defenses rest on underlying assumptions. For example, in the DBA of de Finetti, he implicitly presupposes we agree with him on the following normative statements:
        \begin{itemize}
            \item [-] credences are betting dispositions,
            \item [-] belief guides action,
            \item [-] losing money is bad.
        \end{itemize}
        These are value judgments, which indeed are the roots of a statistical flower: the subjective Bayesian flower. How could we ever refute these value judgments? Ellsberg gives it a shot. 
        
    \section{The paradox of Ellsberg's paradox}
    \label{sec: Paradox}
        The Ellsberg paradox is a thought experiment where agents seem to behave irrationally. It was first introduced by \cite{keynes2013treatise} and then later revisited by \cite{ellsberg1961risk}. This section crudely describes the two-urn paradox. Urn I contains 50 red and 50 black balls, and Urn II contains 100 red and black balls \textit{where the ratio is unknown}. Then, the experimenter offers the following bets:
        \begin{itemize}
            \item [(i)] get 1 util if you draw red from Urn I,
            \item [(ii)] get 1 util if you draw black from Urn I,
            \item [(iii)] get 1 util if you draw red from Urn II,
            \item [(iv)] get 1 util if you draw black from Urn II,
        \end{itemize}
        and 0 otherwise for each bet. A typical participant strictly prefers (i) to (iii) \textit{and} (ii) to (iv). Credences then sum to higher than one, violating the second axiom of Probabilism. Decision-makers with such irrational preferences can be Dutch-Booked. 
        \paragraph{}
        The paradox raises serious doubt about the postulates Savage and de Finetti laid out (see \cite{epstein1993dynamically}).
        On the one hand, we have the \textit{subjective Bayesians} who require agents to adhere to the Probabilism rule, and on the other, \textit{Ellsberg's observation} that demonstrates that agents' credences are systematically irrational. Should we just refute subjective Bayesianism? Or Bayesianism altogether? \\
        No, subjective Bayesianism is a \textit{normative} idea for rational agents, while Ellsberg's observation is a \textit{descriptive} falsification. Unfortunately, these two do not clash. A (subjective) Bayesian can simply say that these participants acted irrational, that is, that they did not follow the prescription. \\
        But why does this matter? The following gives reason, presented as a premise-conclusion form.
        \begin{itemize}
            \item[(P1)] Empirical falsification does not lead to normative failing.
            \item[(P2)] There exists more than 1 normative school for decision theory (henceforth “NSD”), such as Bayesianism, Minimax theory (related to Pareto efficiency), Frequentism, Fiducial probabilities, and Likelihoodism.
            \item[(P3)] A school of statistics is implied by the choice of the NSD.
            \item[(P4)] Researchers ought to choose a normative theory of statistics in their statistical endeavours (such as estimation or hypothesis testing)
            \item[(I1)] By (P2), (P3), and (P4), methodologists need to construct a way to choose between NSDs.
            \item [(C)] By (P1) and (I1), as empirical falsification is not able to eliminate NSD, methodologists should find a different way.
        \end{itemize}
        See the Appendix for defences for each premise.
        This means that, due to the multiplicity of NSDs, we should look for an approach to choose among them. Section \ref{sec: Meta-problem} presents two candidate methods, and shows why one of them falls short. But first, we explain the problem in more detail.

        \subsection{Statistical endeavours and how they ought to be performed}
            When a researcher is interested in a statistical endeavour she needs to pick out a school, which dictates how the endeavour should be performed. If we adopt Frequentism, we believe in a theory that is supported by the idea that we have repeated experiments. For estimation, this means we want to pick consistent, efficient, and asymptotically normal estimators; for inference, we rely on valid significance tests.\footnote{Furthermore, we may find properties such as consistency against the alternative model, and asymptotic size control to be appealing.} \\
            When subscribing to Likelihoodism, we believe that the likelihood function is the \textit{only} sufficient statistic for the data \citep{berger1988likelihood}. For estimation, this means performing maximum likelihood estimation (MLE); for inference, we need to use the likelihood ratio test (LRT). 
            %Note that the protocols implied by Frequentism and Likelihoodism do not necessarily clash.
            \\
            Endorsing subjective Bayesianism, we believe that there are no repeated experiments, that the researcher's expertise (subjective prior) should influence the research conclusion, and that new evidence should be handled using Bayes' rule (the Principle of Conditionalization). For estimation, we would simply deliver the posterior distribution (or a sampler); for inference, we use Bayes factors. \\
            Choosing a school inevitably leads to different protocols, as each school rests on its own distinct set of value judgments. In the case of subjective Bayesianism, the principles are distinctively defended by devices---such as de Finetti's DBA---which themselves presuppose certain underlying value judgments (see Section \ref{subsec: foundations}). While the resulting protocols may sometimes align, this is generally not the case. Each school, in turn, has its proponents who aim to steer you away from the competition. Here is Savage giving it his best:
            %Savage provides a clear example of this:
            \blockquote{\textit{“Fisher's school, with its emphasis on fiducial probability - a bold attempt to make the Bayesian omelet without breaking the Bayesian eggs - may be regarded as an exception to the rule that frequentists leave great latitude for subjective choice in statistical analysis. The minimax theory, too, can be viewed as an attempt to rid analysis almost completely of subjective opinions, though not of subjective value judgments.”} \\ \cite{savage1961foundations}, p. 578}

            To me, Savage is saying that frequentists like to portrait themselves as value-free and objective as employing this school leaves you not a lot of choice when performing statistical endeavours. Estimators should be \textit{efficient} (unbiased and maximally precise), and hypothesis tests should be \textit{valid} and uniformly most powerful; but that does not mean the approach is free of value judgments. Subjective Bayesians, however, make it clear that there are subjective value judgments, and even subjective choice within the method: namely, choosing the prior. Still, this means we are stuck with letting each researcher choose the subjective value judgments she deems normative.
            %value judgements that are not made explicit most of the time.
            
    \section{Meta-problem: choosing between normative systems?}
    \label{sec: Meta-problem}
        If empirical falsification alone is not able to refute a normative system, then by what criteria should a researcher choose her school of statistics?\\
        We provide two candidate approaches: 
        \begin{enumerate}
            \item A \textit{universalist approach}, advocating for a single normative foundation---such as subjective Bayesianism or Likelihoodism---that applies uniformly across all settings. Deviations from this foundation are seen as irrational choices.
            
            \item A \textit{context-dependent approach}, which recommends tailoring the choice of normative system to each research context. Different epistemic contexts or questions may call for distinct schools. For example, Frequentism may be well-suited to large-sample inference, while subjective Bayesianism could be appropriate in cases requiring expert elicitation.
        \end{enumerate}
        I will next explain how the universalist approach could be used to evaluate whether a particular school of statistics serves as a normative foundation and why this approach ultimately falls short. This style of argument is fully inspired by \cite{worsdale2021my}'s \textit{My objectivity is better than yours} paper. The authors explore universalist versus context-dependent objectivity in the search for an objective gender inequality index. In particular, they present the paper from \cite{stoet2019simplified}, who argue that mainstream metrics like the Global Gender Gap Index (GGGI) contain systematic biases. In doing so, \cite{stoet2019simplified} submit their Basic Index of Gender Inequality (BIGI) as free from subjectivity; to be specific they try to eliminate, among other things, the feminist and cultural perspectives that GGGI has. \cite{worsdale2021my} then argue why BIGI \textit{also} contains subjective value judgments, and that context-\textit{in}dependent claims to objectivity are worrying. These types of objectivity are generally criticised by philosophers of science \citep{megill1994rethinking, nagel1989view, reiss2014scientific}. Instead, the focus is now on the \textit{operational} objective from \cite{douglas2004irreducible}. The most important factor in this notion is that to either challenge or enhance the objectivity of something, as well as choosing the best methods for addressing these challenges or achieving those enhancements, are usually shaped by the particular details of the context in which it is meant to be applied. This concept of operational objectivity is what inspires the context-dependent approach to choosing a normative system for statistics. But, before introducing this, let us first look at the universalist approach.
        
        \subsection{The universalist approach to choosing a normative system}
            So how exactly does a universalist choose between two normative foundations for statistics? \\
            The universalist approach holds that statistical practice ought to be grounded in a single, overarching normative framework. According to this view, a researcher commits to one statistical school and consistently applies its warranted methods across all contexts. This commitment is guided by the belief that there exists a single best school of thought---one that must be identified through ongoing methodological debate until it stands alone as the foundation every researcher ought to adopt. This ideal of a final victor has shaped statistical discourse for decades, with methodologists locked in ongoing debate, each declaring, \textit{my statistics is better than yours}. Let us sketch the problem with this approach. \\
            Suppose that we could enumerate all possible statistical endeavours: from performing a one-sided binary hypothesis test to predicting future data points using a non-linear model. Imagine that Likelihoodism is chosen as the universal foundation for statistics. This principle appears rational in a decision-theoretic context (from Savage's or \cite{birnbaum1962foundations}'s postulates), but what if it produces a “silly” type of one-sided binary test, such as a test that never rejects? Or worse, what if such an endeavour is not well-defined under Likelihoodism? As it turns out, this challenge is indeed encountered in composite hypothesis testing, as the original LRT assumes simple hypotheses. This can lead to ambiguities under a strict Likelihoodist framework \citep{berger1988likelihood}.\\
            So how do researchers perform LRTs in a composite hypothesis testing setting? Usually, a \textit{generalised} LRT (GLRT) is warranted. The problem is that the GRLT is not derived under Likelihoodism - we need \textit{Generalised} Likelihoodism (set up in \cite{berger1988likelihood}), which does the trick. Problem solved? \\
            In composite settings, kind of\footnote{See \cite{koning2024continuous} for a recent take on the problem.}, but all of these approaches rely on the existence of a likelihood function, which is not always the case. \\
            A few technicalities. For any given statistical endeavour, a likelihood function is a density on the possible data generation process (DGP). \cite{halmos1949application} show that to be able to define this density, we need to have a dominating reference measure that simultaneously dominates the full model (all possible DGPs). In most statistical settings, such a measure exists and is well-defined, such as in a Gaussian location model. But in more complex settings, we run into problems. For example, we might want to test whether the data is drawn from a continuous distribution against the alternative hypothesis that the data is drawn from a discrete Uniform distribution. For this endeavour, a single dominating reference measure does not exist. This, in turn, means that a likelihood ratio test statistic does not exist, even under generalised Likelihoodism.
            However, recently \cite{larsson2024numeraire} has fixed this issue in the hypothesis testing setting: they define an “effective null hypothesis” for which a dominating reference measure can always be specified. The problem remains unsolved for general estimation endeavours (such as MLE), where it is not clear how one should define the likelihood function as there are no competing hypotheses.\footnote{To be specific, the idea in \cite{larsson2024numeraire} only requires that the alternative density is absolutely continuous with respect to all the densities in the null hypothesis. This is a weak requirement in the context of hypothesis testing because if it were violated, the alternative would assign positive probability to some event that has zero probability under every element of the null; if such an event occurs, the null can be immediately rejected at any level. In the context of estimation, the same story cannot be used.} In the future, such techniques might be developed, but relying on future research does not seem to be an appealing feature of the universalist approach. 
            \\
            A hardcore likelihoodist might then respond by saying that such specific settings should not be considered at all, exactly because they are not well-defined. Very well, but then the multiplicity of schools of statistics remains. This is because each vendor can simply sketch the limitations of their school and say that anything outside of it must be an irrational endeavour, not to be performed in research. The following remark presents the same argument against the universalist approach through a linguistic analogy:
            \begin{remark}[Problem with the universalist approach]
                Consider a researcher seeking to perform statistical endeavours according to a single universal normative foundation for statistics. 
            
                Let us define the elements of the analogy as follows:
                \begin{enumerate}
                    \item \textit{researcher} $\equiv$ individual wanting to communicate,
                    \item \textit{statistical endeavour} $\equiv$ linguistic endeavour,
                    \item \textit{single normative foundation} $\equiv$ single language (e.g. English).
                \end{enumerate}
                
                Suppose English is the universal normative language. The individual wants to write a haiku using a term that conveys both tragedy (negative) and greatness (positive). English lacks a single word to capture this duality, while other languages, such as French, might offer a fitting word, like “terrible,” which conveys both senses.
            
                Now, should the individual switch to French? Perhaps, but this transition means losing certain unique aspects of English. The critical question then becomes: does a rational universal language require such a term for all \textit{rational} linguistic endeavours? A universalist might argue that needing such a term signals an irrational preference and so being deviant. But in that case, the challenge of multiple normative foundations remains unresolved.
            \end{remark}
    
            This shows that the universalist approach, which might seem like the only \textit{objective} approach, ultimately cannot resolve the multiplicity of normative systems. The researcher is still left with an arbitrary choice.
    
            Next, we turn to the second candidate: the context-dependent approach. 
    
        \subsection{The context-dependent approach to choosing a normative system}
            The context-dependent approach does not consider a single normative school for statistics but rather posits that different research contexts require specific normative schools. If we are in a setting where data is abundant, yet expert opinion is not clear-cut, subjective Bayesianism does not seem to (bene)fit the context. Frequentism might be more appropriate. This context-dependent protocol also encourages a more thoughtful selection of the research methodology. Namely, if a researcher needs to make a choice between normative schools of statistics, she needs to be more aware of the value judgments she is endorsing. This adds transparency to the research, and encourages a careful alignment between methodology and research goals. 
            \\
            We want to stress that agents can still be irrational under the context-dependent approach. For example, if the research context fits the likelihoodist school, statistical endeavors should be performed using likelihoodist protocols.   
            To illustrate how this context-dependent approach can be used practically, we now turn to a case study.
            
        \subsection{Choosing a normative system: a small case study}
            We will look at the recent paper from \cite{cordes2024motivated} called \textit{Motivated Procrastination}. The authors investigate why people sometimes delay tasks despite understanding the costs. Rather than simply attributing procrastination to impulsive preferences, they explore how people may intentionally hold overly optimistic beliefs about their future effort, which leads them to defer work. They find that when individuals have more room for motivated reasoning, they tend to believe tasks will be easier, thus pushing the work to a later time. \\

            Participants engaged in a four-week longitudinal experiment where they completed an unpleasant task (transcribing numbers) by a deadline. However, the actual workload required was uncertain, giving room for belief-based procrastination. The participants were able to start the workload in the first session, knowing they would have two more sessions later. Then, to model these beliefs, the authors use the subjective Bayesian school, by eliciting the subjective priors on the future work. This was done  during the first week of the experiment. After the prior elicitation, participants were hit by a noisy, yet informative signal on the workload. Specifically, the subjects knew that they were placed in a group and that each person's workload was randomly drawn from a discrete distribution (without replication).  The signal then was how many of the three other workloads (assigned to others) were higher than their own. If, for example, a participant learns that all three other workloads were higher than their own, she updates her prior, forming a positive expectation on the future workload. Then, the authors tactically wait for two weeks before the second session starts. During this session, a participant is either reminded (treated) or not (control) of the signal they received two weeks prior, after which they elicit the personal posterior of each subject. The authors posit that a rational perfect-memory agent must form their posteriors, using Bayes' rule, alluding to the second Bayesian norm of conditionalisation. This means that the control group can have two types of biases when forming posterior beliefs: imperfect memory and irrational updating. The treated group can only display the irrational updating bias. Differences across these two groups identify the (causal) effect of memory on beliefs updating. In turn, \textit{motivated memory} can be identified if the control subjects choose to suppress the negative news more than the positive news, \textit{compared} to treated individuals (difference-in-difference style). 

            To test the hypothesis of motivated memory, they employ frequentist tests: mostly two-sided $t$-tests.\footnote{Using their dataset of 367 observations, they perform around 120 $t$-tests. Bonferonni could probably not sleep at night hearing this. This is an example where the authors do not follow the frequentist protocol.} But, their model for a rational individual is Bayesian; how can this be warranted? Though I could not imagine the authors were thinking of this, they engaged with the context-dependent approach. For one part of their method, they modelled the decision using the Bayesian idea - for rational belief, that is - while for the other, they used the Frequentist strategy - namely, performing significance tests. Now the question is not whether it is warranted to use these schools side-by-side, but rather, whether the employed school is warranted \textit{in each respective context}. To use Bayesian to represent an agent who i) elicits a prior, ii) receives evidence, and iii) updates and reveals her posterior seems to be appropriate. Modelling rational agents as subjective Bayesians is customary in game theory and experimental economics, and for good reason: its clearly defined decision-theoretic roots. Now for the hypothesis testing. Testing is also about a decision - to reject or not to reject - so why switch to Frequentist tests? Bayes factors could indeed have been used to test whether the alternative hypothesis (motivated memory) has stronger evidence compared to the null hypothesis (no motivated memory). But, as the researchers had no particular prior on the hypothesis, they could not benefit from a subjective Bayesian approach. Frequentism seems to be an appropriate solution. We want to note that the researchers could have used \textit{objective} Bayesianism for the testing part of the research. In this school, the prior is not related to the knowledge of the analysts and instead uses the information of the model; in this case it was a Gaussian location model. \\
            
            The point of this case study is to show that one school can be appropriate under a specific context, while less so on another. This case study had two separate contexts: one for modelling rational beliefs and updating, and one for hypothesis testing. Subjective Bayesianism seemed appropriate for the former, while not for the latter. 
            
    \section{Conclusion}
    \label{sec: Conclusion}
        This essay tries to look at a universalist versus a context-dependent approach to choosing statistical norms. The former endorses a single universally applicable normative foundation for statistics, while the latter posits that the choice of the school of statistics should be matched with the research context. I advocate for the context-dependent approach, as it offers a way to cater to the research questions and goals at hand, instead of obsessing over which statistical school ought to be chosen once-and-for-all. The inspiration comes directly from \cite{worsdale2021my} who also support context-driven notions of objectivity from \cite{douglas2004irreducible}. They apply the idea to the search for an objective measure of gender inequality. \\
    
        The context-dependent approach I have outlined above is of course not free from obscurities. Firstly, it is not entirely clear what to do when the context is ambiguous. Let us suppose that we have an expert in our research team, but it is not clear whether he truly is mastering the subject. Should we elicit his prior and perform a Bayesian analysis? Or, in a setting with 100 replications — exceeding the typically small sample sizes in experiments like western blotting — does this justify adopting a Frequentist approach? To answer these operational questions, we need to consult the statistics books, which usually come up with rules-of-thumb that are context-specific. \\
        And what if we only have minimal context, such as two data points and no prior? Many statisticians would recommend for you to simply “look at the data and draw your own conclusions”, instead of forcing the two data points into a testing procedure (as many of its assumptions may not hold). But looking at the data and drawing your own conclusions is a form of subjective Bayesianism, where computation and testing happen in the neural networks of the researcher. These cases should be considered carefully, which is exactly what could complement this essay. \\
    
        Secondly, within a school, such as Frequentism, there are, just as in the Bayesian school, divisions. In hypothesis testing, controlling the false positive rate at a fixed level (validity), and minimising the false negative rate (maximising power), is a Neyman-Pearson approach (with emphasis on so-called inductive behavior). These significance tests result in a binary “rejection or not” decision. On the other hand, there is the school of Ronald Fisher, which does not study these two types of error and directly considers level $p$ tests for given $p$-values - something which would not be \textit{valid} under strict Neyman-Pearson Frequentism. 
        Finally, the context-dependent approach does not remedy instances where a researcher makes a conceptual mistake when performing a test under a specific (sub)school of statistics, such as performing \textit{multiple} tests for a single dataset without controlling for a blown-up false positive rate \citep{benjamini2002john}. It is still up to the researcher to figure out the exact protocols that are prescribed by the chosen school. \\
        
        An appealing artefact that could come from the context-dependent approach is that as researchers need to be more aware of their methodology, they will more carefully employ their statistical procedure. In turn, matters such as the multiple testing problem become more apparent. 
    
    \bibliographystyle{apacite}
    \bibliography{References}
    \newpage
    \section{Biographical note}
        Dear reader,\\
        I am currently a research master student at the Tinbergen Institute in Amsterdam. Before starting my graduate studies, I finished a bachelor in econometrics at the EUR. I became highly intrigued in the of field of statistics during that time, and my curiosity has only grown. At the time of writing this essay, I am the most interested in the various decision-theoretic foundations of statistics. Usually, these roots lead to Bayesianism and Likelihoodism. At the other side of the foundation of statistics, lies (mostly) measure theory. These roots tend to favour Frequentism. I have also come to learn that the (sub)schools are highly ambiguous.\footnote{Within Bayesianism, there is a variety of subscriptions you can choose from, such as pure, subjective, objective, or Frequentist (yes!) Bayesianism (see \cite{kleijn2020frequentist} for an in-depth discussion). Within Frequentist hypothesis testing, as I briefly point out in the essay, there are Neyman-Pearson and Fisher approaches. These are not directly related to Fisher's fiducial probability system, \textit{or} the famous Neyman-Pearson Lemma. If anything, the latter is rather a Likelihoodist approach.} Though both foundations seem to be irrefutable in their own respect, they can lead to different statistical prescriptions. So, if mathematics (deductive) is the only medium through which the roots express themselves, which clash, then it must be that they try to answer fundamentally different questions. Should we instead look for a statistical school that can cover any foundations? \\
        
        I am only at the beginning of comprehending these questions, and with the modern \textit{e-values} --- which are currently causing a renaissance in testing \citep{ramdas2024hypothesis} --- it is becoming even more difficult. Currently, I am learning measure theory to improve my intuition on the subject. Until now, my intuition is telling me that defending one specific normative school of statistics is an obsessive ideal (like a religion), and that something such as a context-dependent normative system seems more adequate. The philosophy teachers I have talked to have generally liked the idea, while the economics teachers I have talked to prefer to religiously stick to their school (econometricians tend to be Frequentist, while decision-theoretic economists are Bayesians). Who should I trust? \\

        I write this essay in order to organise my thoughts about the meta-problem I sketch. Hopefully, the approach I propose gets refined or refuted, so that I can come to a deeper understanding about the subject. 
    \section*{Appendix}
    \label{sec: Appendix}
    \appendix
    \section{Premise defence for Section \ref{sec: Paradox}}
        (P1) is can be defended as such: A normative statement is one that prescribes how one \textit{ought} to act. If one does not act as the prescription (empirical falsification), they simply did not follow the prescription, there is no need to change the normative statement. Note that this defence seems to go against the context-dependent approach I am proposing. It does not. Suppose we are adopting the context-dependent approach. What I am then trying to say is that for a given context, a certain school is implied. This school should then be followed normatively. Suppose that the context alludes to subjective Bayesianism, then i) one ought to form priors that are probability measures, and ii) one ought to use Bayes rule for updating. Any one who deviates within this context is irrational, even in the seemingly flexible context-dependent approach. In the universalist approach, the original defence works by itself.\\

        \noindent (P2) is true as alternatives to subjective Bayesianism exist, though these alternatives can be related. One that is quite disjoint from subjective Bayesianism is Frequentism (see in Section 3.1) or Minimax theory. \\

        \noindent(P3) is too difficult to defend in this essay. \cite{savage1972foundations} uses this exact idea to go from decision-theoretic postulates to subjective Bayesianism. \cite{wald1947foundations} too gives decision-theoretic foundations of Bayesianism, proving that Bayesian procedures are the only admissible ones. For Frequentism, \cite{lehmann1986testing} describe that much of its theory is based on the Minimax theory; a concept closely related to Pareto optimal decisions. Fiducial probabilities are also defended using Ronald Fisher's decision-theoretic ideas. \\
        
        \noindent(P4) is a construction for science. If researchers come up with their own ways to draw conclusions from data, it would become difficult to communicate ideas. Of course, a skeptic might be against the use of data in science altogether, and only endorse the use of \textit{logic} for scientific progress. There is a problem with this. The fundamental reason scientists can entertain themselves is that they are ignorant one some statements (assumption). Their job is to come up with some expert's opinion about the quality of the hypothesis in question. Logic can only model truth or false statements, which cannot accommodate credences. Fuzzy logic can model \textit{vagueness} in the logical answer, but not \textit{ignorance} on the statement \citep{asli2017handbook}. That is what probabilities are for, which are the mathematical roots of statistics. There are non-probabilistic models for statistics, such as Kolmogrov's structure function or Shafer's game-theoretic approach. The former uses notions of entropy, while the latter employs stochastic dominance, yet both are data-driven. Greek logic is simply not enough in the scientific enterprise. In the Dutch code of conduct for scientific integrity, researchers are obliged to use warranted methodology, which includes statistical analysis, so I hope this defence is convincing enough.
\end{document}